\documentclass[10pt,conference]{IEEEtran}
\IEEEoverridecommandlockouts
\usepackage{cite}
\usepackage[hidelinks]{hyperref}
\usepackage{amsmath,amssymb,amsfonts}
\usepackage{algorithmic}
\usepackage{graphicx}
\usepackage{textcomp}
\usepackage{xcolor}
\def\BibTeX{{\rm B\kern-.05em{\sc i\kern-.025em b}\kern-.08em
    T\kern-.1667em\lower.7ex\hbox{E}\kern-.125emX}}
\usepackage{comment}
\usepackage{url}
\usepackage{amsthm} % for definitions
\usepackage{mathtools} % to put def word above equals sign
\usepackage{listings}
\usepackage{subcaption} % For captioning subfigures
\usepackage{natbib} % for citet

\theoremstyle{definition}
\newtheorem{definition}{Definition}[section]

\newcommand\eqdef{\stackrel{\mathclap{\normalfont\mbox{\scriptsize{def}}}}{=}}

\begin{document}

% Used title from 2019 CASCON Poster
\title{A Web App for Teaching Finite State Automata \\
}

\author{\IEEEauthorblockN{Christopher William Schankula}
\IEEEauthorblockA{\textit{Department of Computing \& Software} \\
\textit{McMaster University}\\
Hamilton, Ontario, Canada \\
schankuc@mcmaster.ca}
\and
\IEEEauthorblockN{Lucas Dutton}
\IEEEauthorblockA{\textit{Department of Computing \& Software} \\
\textit{McMaster University}\\
Hamilton, Ontario, Canada \\
duttonl@mcmaster.ca}
}

\maketitle

\begin{abstract}
We present the open-source tool \texttt{finsm.io}, a tool for creating, simulating and
exporting deterministic and non-deterministic finite state automata (DFA/NFA). We first
describe the conceptual background on which the tool is based, followed by a description of
features and preliminary evaluation of the tool based on use spanning multiple years and 
hundreds of student users. Preliminary evaluation found that instructors and students 
overwhelmingly recommend the tool to others and agree that it has improved their learning 
and teaching. The authors invite interested educators to use the tool in their finite
automata courses.
% This document is a model and instructions for \LaTeX.
% This and the IEEEtran.cls file define the components of your paper [title, text, heads, etc.]. *CRITICAL: Do Not Use Symbols, Special Characters, Footnotes, 
% or Math in Paper Title or Abstract.
\end{abstract}


\section{Introduction} % (Chris)
We present the open-source tool \texttt{finsm.io}\footnote{\url{https://finsm.io}, source code: \url{https://github.com/CSchank/finsm/}}
which has been successfully used to teach
finite automata courses in software engineering and computer science courses spanning
multiple years and 860+ students. The system is accessible as a free web 
application, and allows users to build, simulate (test) and export finite state automata.

In this tool paper, we describe the theory and background of the system, detail its 
implementation and features available, then present a preliminary evaluation based on 
written responses and surveys with instructors, teaching assistants, and students. Finally, we
present prior research in this area and conclude with ideas for future research and 
improvements.

Students and instructors alike overwhelmingly agreed that they would recommend the system 
and that it benefitted learning and teaching. The authors are happy to offer the tool
to any like-minded instructors and encourage anyone who is interested to contact us.

\section{Background}\label{sect:background}

We follow the definitions given in \cite{kozen2012automata} for finite
state machines. 

\begin{definition} % p.15 textbook
\label{def:dfa}
A \textit{deterministic finite automaton} (DFA) is a 5-tuple
\vspace{-5mm}
\begin{align*}
M = (Q, \Sigma, \delta, s, F)
\end{align*}

where

\begin{itemize}
    \item $Q$ is a finite set; its elements are called \textit{states}.
    \item $\Sigma$ is a finite set; called the \textit{input alphabet}.
    \item $\delta : Q \times \Sigma \rightarrow Q$ is the \textit{transition function}.
    \item $s \in Q$ is the \textit{start state}.
    \item $F \subseteq Q$; its elements are called \textit{accept/final} states.
\end{itemize}

\end{definition}

\begin{definition} % p.32 textbook
\label{def:nfa}
A \textit{non-deterministic finite automaton} (NFA) is a 5-tuple
\vspace{-3mm}
\begin{align*}
M = (Q, \Sigma, \Delta, S, F)
\end{align*}

where $Q$, $\Sigma$ and $F$ are the same as in Definition~\ref{def:dfa}, and

\begin{itemize}
    \item $S \subseteq Q$ is the set of \textit{start states}.
    \item $\Delta : Q \times \Sigma \rightarrow \mathcal{P}(Q)$ is the transition function returning the \textit{power set} of $Q$.
\end{itemize}
\end{definition}

Given that every DFA $(Q, \Sigma, \delta, s, F)$ is equivalent to an NFA
$(Q, \Sigma, \Delta, \{s\}, F)$, where $\Delta(p, a) \eqdef \{\delta(p,a)\}$,
the Elm implementation leverages this theoretical fact to store both machines
using the NFA definition. Hence, Definition~\ref{def:nfa} is represented
as a record type with labeled fields as follows:

\begin{lstlisting}[language=haskell]
type alias Machine =
    { q : Set StateID
    , delta : Delta
    , start : Set StateID
    , final : Set StateID
    , stateNames : Dict StateID String
    ... -- more fields related to layout and names
    }

type alias Delta = 
  Dict StateID (Dict TransitionID StateID)
type alias StateID = Int
type alias TransitionID = Int
\end{lstlisting}

The full definition is elided; only the fields relevant to
the mathematical definition are shown. We treat the
alphabet $\Sigma$ as the set of all symbols that appear
in the transition labels of the machine, and do not store it separately as a set. We make use of the Elm
\texttt{core} library's implementation of 
\lstinline|Set| and \lstinline|Dict| their
use loosely corresponds to the mathematical notion
of sets and (partial) functions respectively.

\subsection{Example}
\begin{figure}[!ht]
    \centering
    \includegraphics[width=0.3\textwidth]{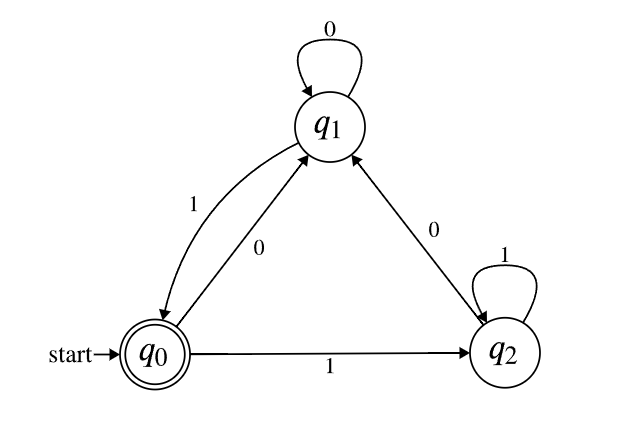}
    \caption{An example DFA created from \texttt{finsm.io.}}
    \label{fig:dfaexample}
\end{figure}
\begin{figure}[t]
    \centering
    \includegraphics[width=0.5\textwidth]{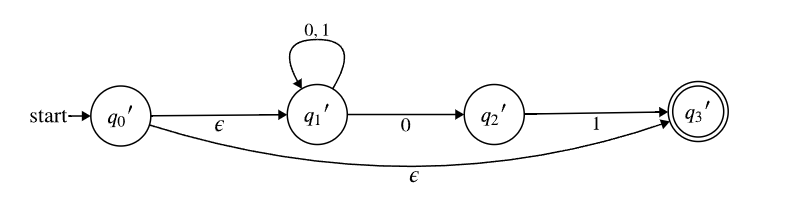}
    \caption{An example NFA created from \texttt{finsm.io.}}
    \label{fig:nfaexample}
\end{figure}

Throughout this paper, we will use the running examples in Figures~\ref{fig:dfaexample}
and \ref{fig:nfaexample}. Both machines accept the string containing `0's and `1's, with the restriction that the string must terminate with ``01''. They also
accept the empty string. Note that while the machines are not the same type, they are \textit{equivalent} in that they accept the same \textit{language} of strings.

Using the mathematical definitions in Section~\ref{sect:background},
Figure~\ref{fig:dfaexample} can be represented as a DFA with 
$Q = \{q_0, q_1, q_2\}$,
$\Sigma = \{0, 1\}$, $s = q_0$, $F = \{q_0\}$, and the transition function:
\begin{align*}
    \delta(q_0, 0) &= \delta(q_1, 0) = \delta(q_2, 0) = q_1 \\
    \delta(q_1, 1) &= q_0 \\
    \delta(q_2, 1) &= q_2
\end{align*}

Similarly, Figure~\ref{fig:nfaexample} can be represented as a NFA with
$Q = \{{q_0}', {q_1}', {q_2}', {q_3}'\}$,
$\Sigma = \{0, 1\}$, $s = \{{q_0}'\}$, $F = \{{q_3}'\}$, and the transition function:
\begin{align*}
    \Delta({q_0}', \epsilon) &= \{{q_1}', {q_3}'\} \\
    \Delta({q_1}', 0) &= \{{q_1}', {q_2}'\} \\
    \Delta({q_1}', 1) &= \{{q_1}'\} \\
    \Delta({q_2}', 1) &= \{{q_3}'\} \\
    \text{otherwise} &= \{\}
\end{align*}

Taking the NFA example further, this is stored in \texttt{finsm.io} as the following Elm expression:

%\begin{minted}[fontsize=\footnotesize,escapeinside=||,mathescape=true]{elm}
\begin{lstlisting}[language=haskell]
nfaExample =
  { q = Set.fromList ["q_0'","q_1'","q_2'","q_3'"]
  , delta = Dict.fromList 
      [("q_0'",Dict.fromList 
        [(Set.fromList ["\epsilon"],"q_1'")
        ,(Set.fromList ["\epsilon"],"q_3'")])
      ,("q_1'",Dict.fromList 
        [(Set.fromList ["0"],"q_2'")
        ,(Set.fromList ["0", "1"],"q_1'")])
      ,("q_2'",Dict.fromList 
        [(Set.fromList ["1"],"q_3'")])
      ,("q_3'",Dict.fromList [])]
  , start = Set.fromList ["q_0'"]
  , final = Set.fromList ["q_3'"]
  , stateNames = 
      Dict.fromList [(1,"q_0'"),(2,"q_1'"),
                     (3,"q_2'"),(4,"q_3'")]
  }
\end{lstlisting}

For clarity, we have substituted \lstinline|StateID|s and
\lstinline|TransitionID|s with their actual values in the
above code snippet. The actual Elm representation uses the
integer IDs, and we can fetch their actual values from other
dictionaries stored in the record when needed.

\section{Features}
\texttt{finsm.io} has 3 main features: 
Building state machines, simulating
them with provided inputs, and exporting to \LaTeX. 
We will illustrate and explain
these features using Figure~\ref{fig:finsmfigs},
which shows the application
in different modes.

\subsection{Build}
\begin{figure*}[h!]
\centering
\begin{subfigure}[b]{.45\linewidth}
\includegraphics[width=\linewidth]{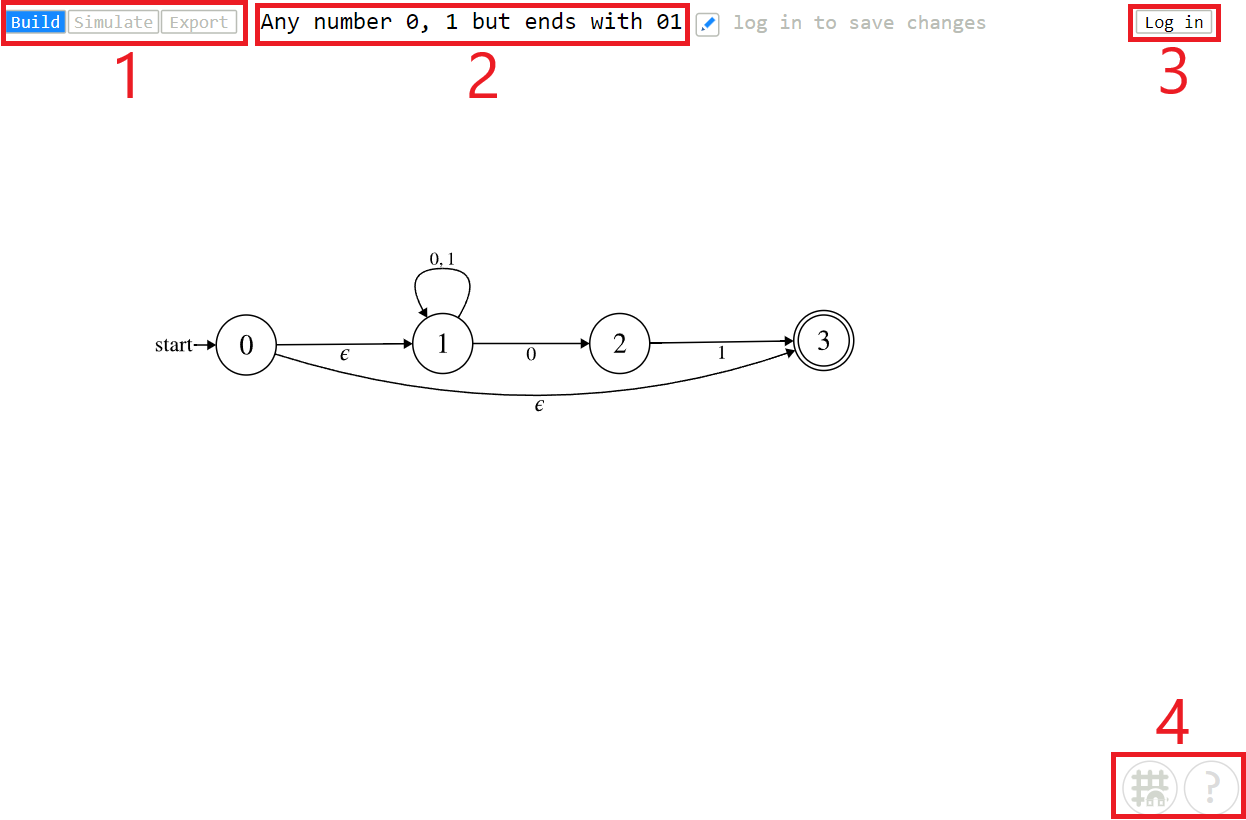}
\caption{The build mode screen. 1. Allows users to switch between the 3 main modes of \texttt{finsm.io}. 2. Users can name their machine here. 3. Users can log in with a Google account, allowing save and load. 4. Buttons for grid-locking and opening a wiki help page.}\label{fig:build}
\end{subfigure}
\begin{subfigure}[b]{.45\linewidth}
\includegraphics[width=\linewidth]{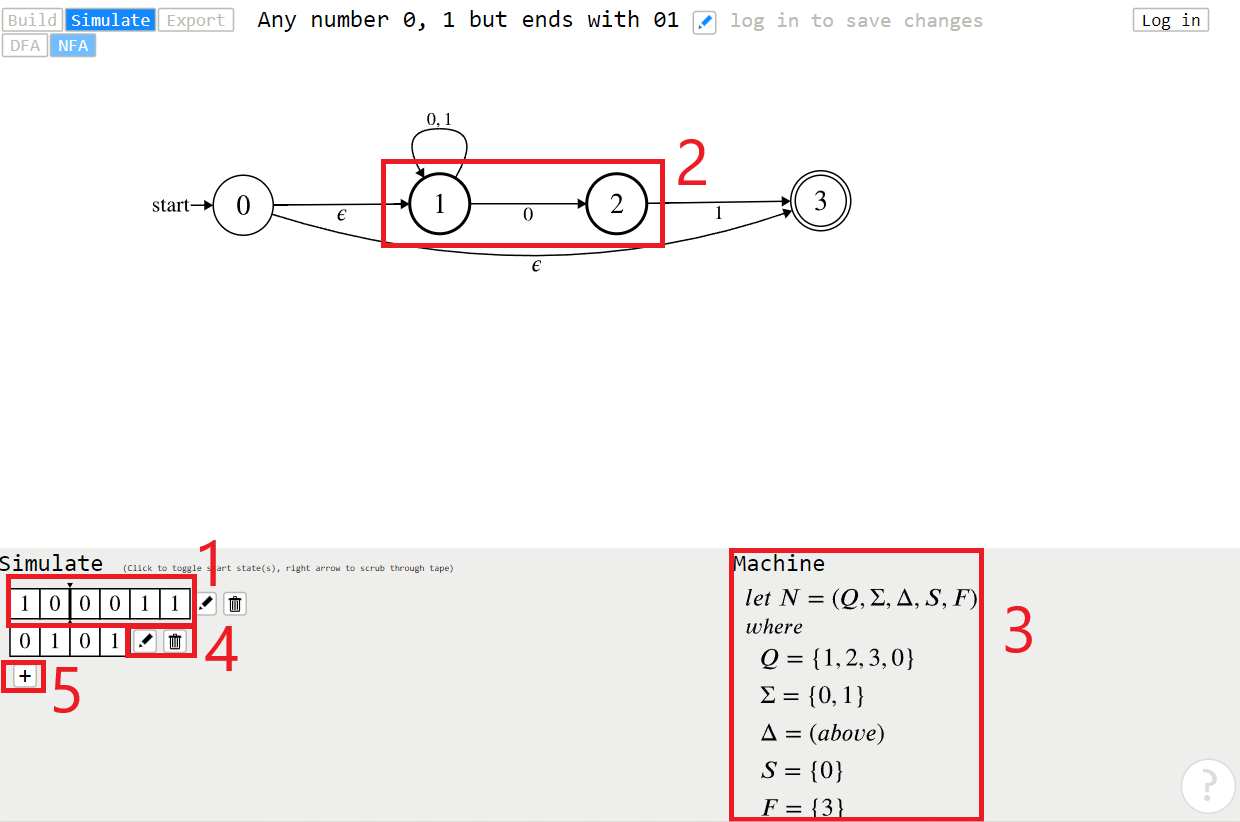}
\caption{The simulation mode screen. 1. The ``tape'' that users can create and run as test input.  2. The machine updates its state as users ``scrub'' through the tape. 3. A mathematical definition of the current machine. 4. Buttons to edit and delete the tape 5. A button to create a new tape.}\label{fig:simul}
\end{subfigure}

\begin{subfigure}[b]{.45\linewidth}
\includegraphics[width=\linewidth]{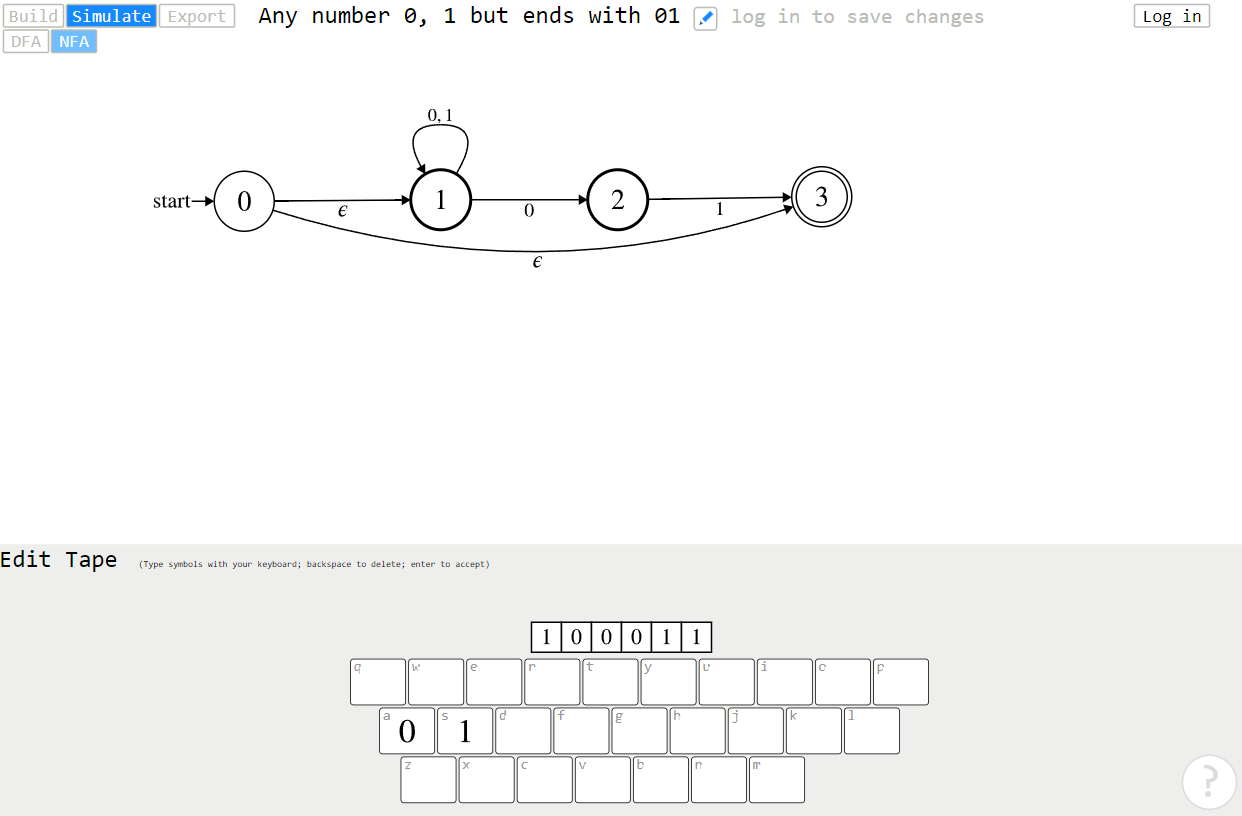}
\caption{Quick keyboard for editing tapes in simulation mode.}\label{fig:simulkeyboard}
\end{subfigure}
\begin{subfigure}[b]{.45\linewidth}
\includegraphics[width=\linewidth]{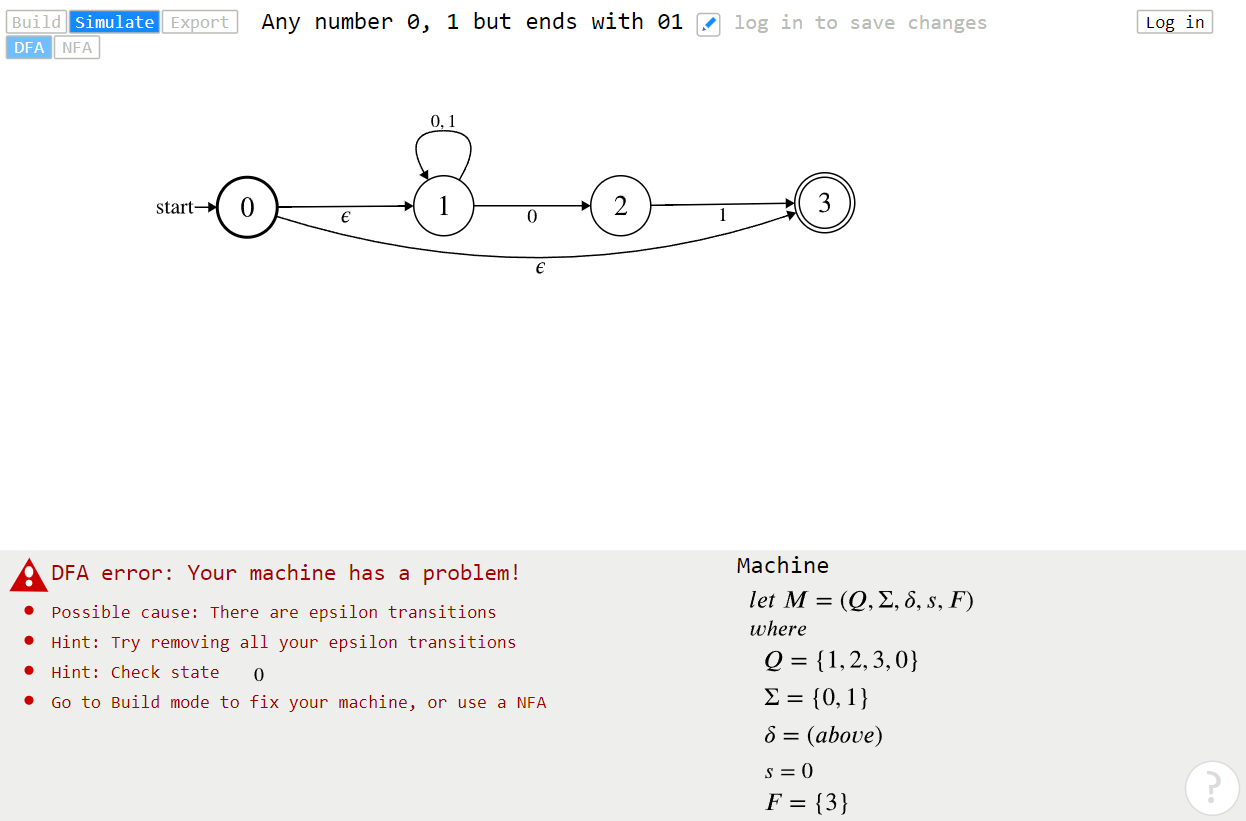}
\caption{A simulation error when the wrong machine type is selected. }\label{fig:simulerr}
\end{subfigure}

\begin{subfigure}[b]{.45\linewidth}
\includegraphics[width=\linewidth]{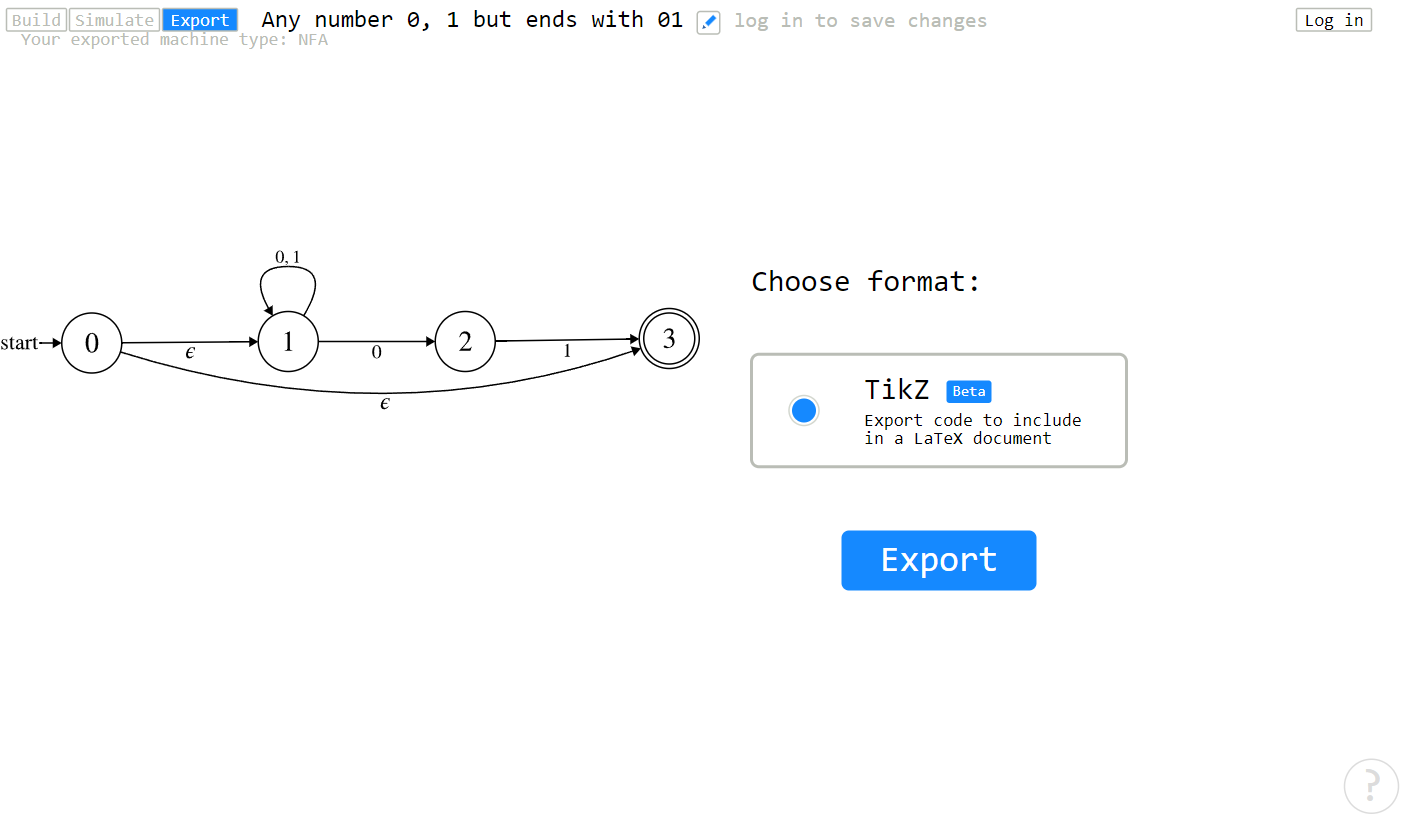}
\caption{The export mode screen.}\label{fig:export}
\end{subfigure}
\begin{subfigure}[b]{.45\linewidth}
\includegraphics[width=\linewidth]{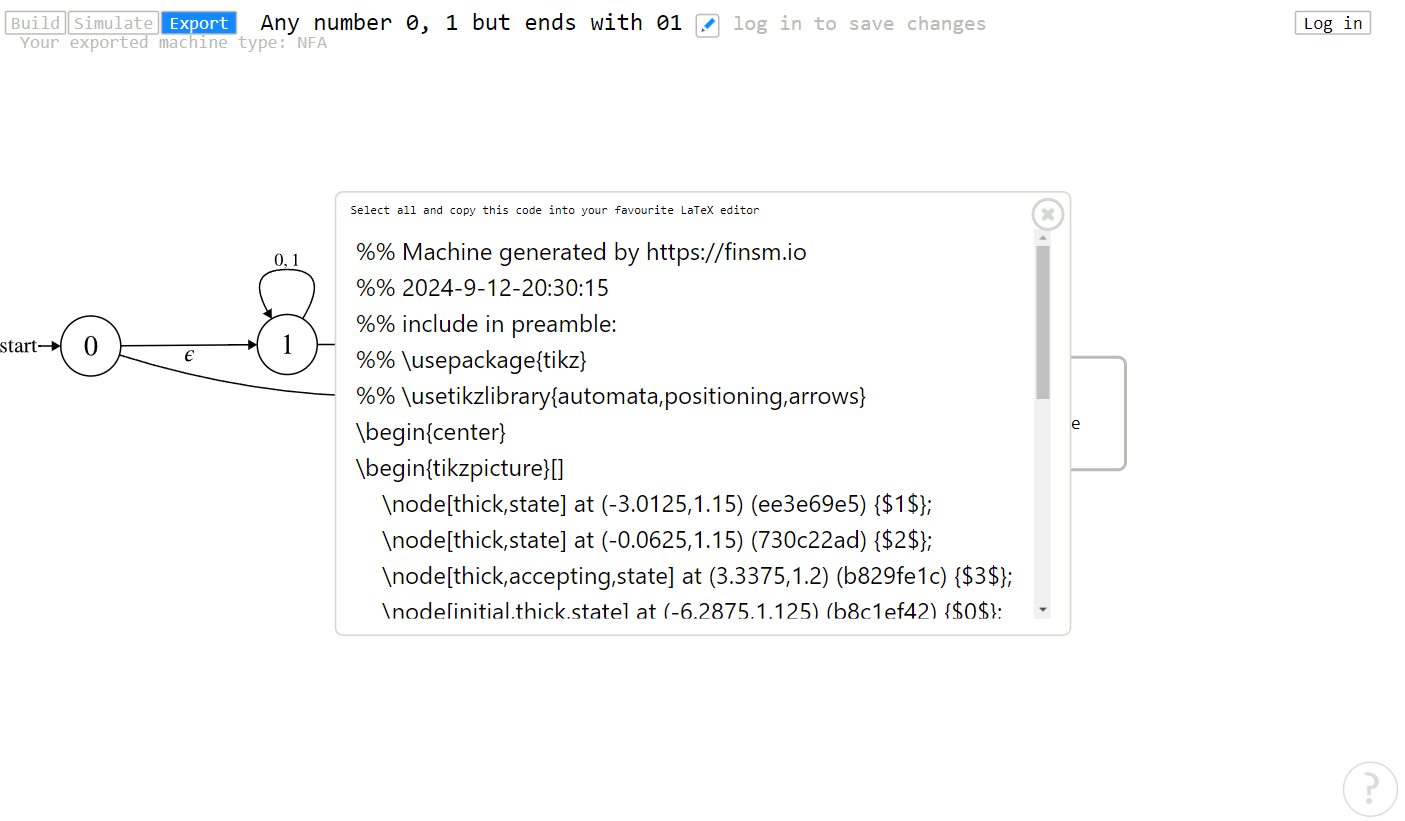}
\caption{A popup containing code that can be copied into a \LaTeX\ document.}\label{fig:exportlatex}
\end{subfigure}
\caption{A collection of screenshots displaying different features of the \texttt{finsm.io} application in a web browser.}
\label{fig:finsmfigs}
\end{figure*}

Figure~\ref{fig:build} lets users construct state machines and is
also the first screen that the users interact with. It is designed
to be minimalistic, only allowing pre-defined keyboard shortcuts
and mouse clicks to create states and transitions.

\begin{figure}[t]
\centering
\begin{subfigure}[b]{.3\linewidth}
\vspace{-5mm}
\includegraphics[width=\linewidth]{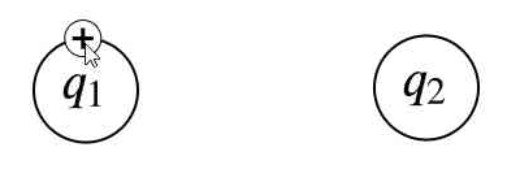}
\vspace{-5mm}
\caption{}
\end{subfigure}
\hspace{2mm}
\begin{subfigure}[b]{.3\linewidth}
\includegraphics[width=\linewidth]{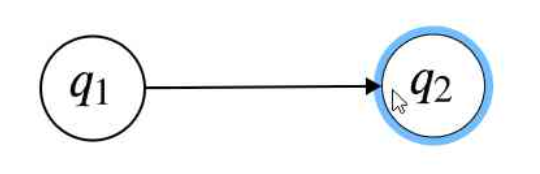}
\caption{}
\end{subfigure}
\hspace{2mm}
\begin{subfigure}[b]{.3\linewidth}
\includegraphics[width=\linewidth]{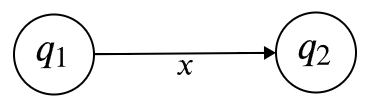}
\caption{}
\end{subfigure}
\caption{A step-by-step illustration on creating transition arrows
in Build mode.}
\label{fig:transitionarrow}
\end{figure}

The typical workflow in this mode involves three actions:
creating new states, creating new transitions, and naming
states and transitions. Creating new states is easily done
by \texttt{Shift}+left-clicking on an empty spot. 
Creating new transitions is shown in Figure~\ref{fig:transitionarrow};
(a) mouse hover over the state to start the arrow from, 
(b) when a visual indicator appears, drag the arrow to the
state to connect to,
(c) release the mouse button to complete the arrow.
Naming is done by \texttt{Shift}+left-clicking a label, which
brings up a text box of the label. \texttt{finsm.io} supports
\LaTeX\ and will render appropriately when users press
\texttt{Enter} after editing the text.

\subsection{Simulate}
Once a user is ready to test their constructed machine,
they can switch over to simulation mode, show in Figure~\ref{fig:simul}. The main feature
is the input tapes on the bottom-left of the screen.
Users can click on a tape and press
\texttt{Right Arrow} to advance the ticker on the
tape. The ticker indicates the current read position
on the tape, and when it is advanced by the user,
the machine will react to the new input by highlighting
states that it transitions into in bold.

When a user edits their tape, it brings up the keyboard
menu in Figure~\ref{fig:simulkeyboard}. \texttt{finsm.io}
automatically detects the alphabet of the language, and
maps a key to each letter. For example, if the user presses
\texttt{a} on their keyboard, it adds a \texttt{0} to the
end of the input tape. Pressing \texttt{Del} and \texttt{Enter}
will delete the rightmost character on the tape and
accept changes made to the tape respectively.

\texttt{finsm.io} will detect errors in constructed machines,
particularly for DFAs. If a user attempts to simulate
the current example as a DFA, they will see the error in
Figure~\ref{fig:simulerr}. Here, they are informed that
$\epsilon$-transitions exist at State 0, which are not
allowed for DFAs. There are other errors as well, 
such as State 2 not having a transition for 0, but \texttt{finsm.io}
will only report one error at a time.

\subsection{Export}
\texttt{finsm.io} supports exporting to \LaTeX\  with the
export mode. It generates \texttt{TikZ} code that can
be copied and pasted into any \LaTeX\ document that
imports the \texttt{TikZ} package. Figure~\ref{fig:export}
displays the main screen, and Figure~\ref{fig:exportlatex}
shows a popup window containing the \LaTeX\ code. The
states in the code are \textit{hashed} and are uniquely generated
whenever a request to export takes place.

\section{Evaluation}

To evaluate the tool's impact on learning and teaching, we asked the instructors,
TAs and students for written and survey-based feedback. This section details some of those 
preliminary answers and other general statistics.

The tool has been used in several finite automata courses in the past five years. A total
of \textbf{868 users} are registered for the website, and this does not include users from 
the first year when there was no login or saving system. These users have used the tool to 
create over \textbf{4,100 state machines} (again, not including unsaved machines from unregistered users).

\subsection{Instructor Perspective}
One of two  instructors responded to a request for feedback. 

\begin{enumerate}
    \item \textit{How did students submit assignments prior to using the tool?}\\
    Students used general-purpose drawing tools or
    even drew machines by hand.
    \item \textit{What are your overall impressions of the tool?}\\
    It is useful for the students and helps the Teaching 
    Assistants (TAs) with marking since assignments are much cleaner.
    \item \textit{What effect did you observe on the students/TAs using the tool?}\\
    Students spent less time drawing automata, and it made debugging and editing easier.
    \item \textit{Do you have suggestions to improve the tool?}\\
    One instructor suggested the addition of Push-Down Automata (PDA), and some 
    visualizations of common finite state automata algorithms.
\end{enumerate}

\subsection{Teaching Assistant Perspective}
Next, we wanted to understand the TA perspective, since they are likely to be the ones
most affected in terms of day-to-day instruction while using the tool.  
Two TAs responded to requests for comment and both ethusiastically recommend the tool.
Notable was the comment that student requests for remarking were reduced because 
the \LaTeX\ diagrams were unambiguous, so easier to read,
and the tool helped students avoid missing transitions.
TAs also used the tool to create assignment questions, which saved a lot of time,
and produced consistently readable results.
One TA requested support for PDAs and both requested support for Turing Machines.

\subsection{Student Perspective}

Anecdotally, students enjoy using the tool and we have observed many users continuing to 
use the tool in future courses where it is not the suggested tool. To quantify this
sentiment, users were sent an optional survey via email.

\begin{figure}
    \centering
    \includegraphics[width=0.8\linewidth]{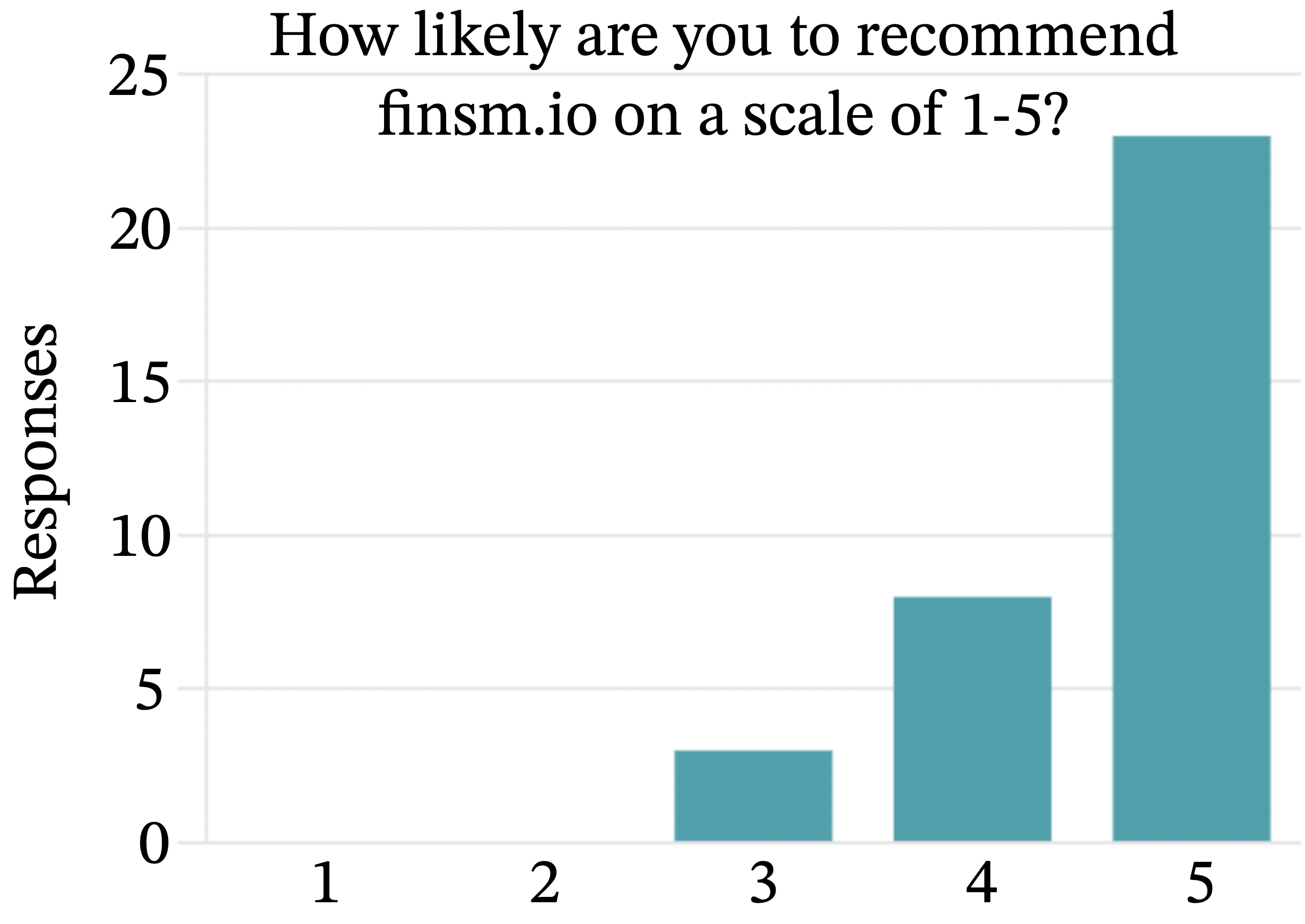}
    \caption{Histogram of student responses to peer recommendation question. 
    [n = 34]. The mean response was 4.59/5.}
    \label{Fig:Q2}
\end{figure}

We received 34 responses. On average, \texttt{finsm.io} received 4.59/5 in terms of recommendation (see Figure~\ref{Fig:Q2}).

\begin{figure*}[h!]
    \centering
    \includegraphics[width=\textwidth]{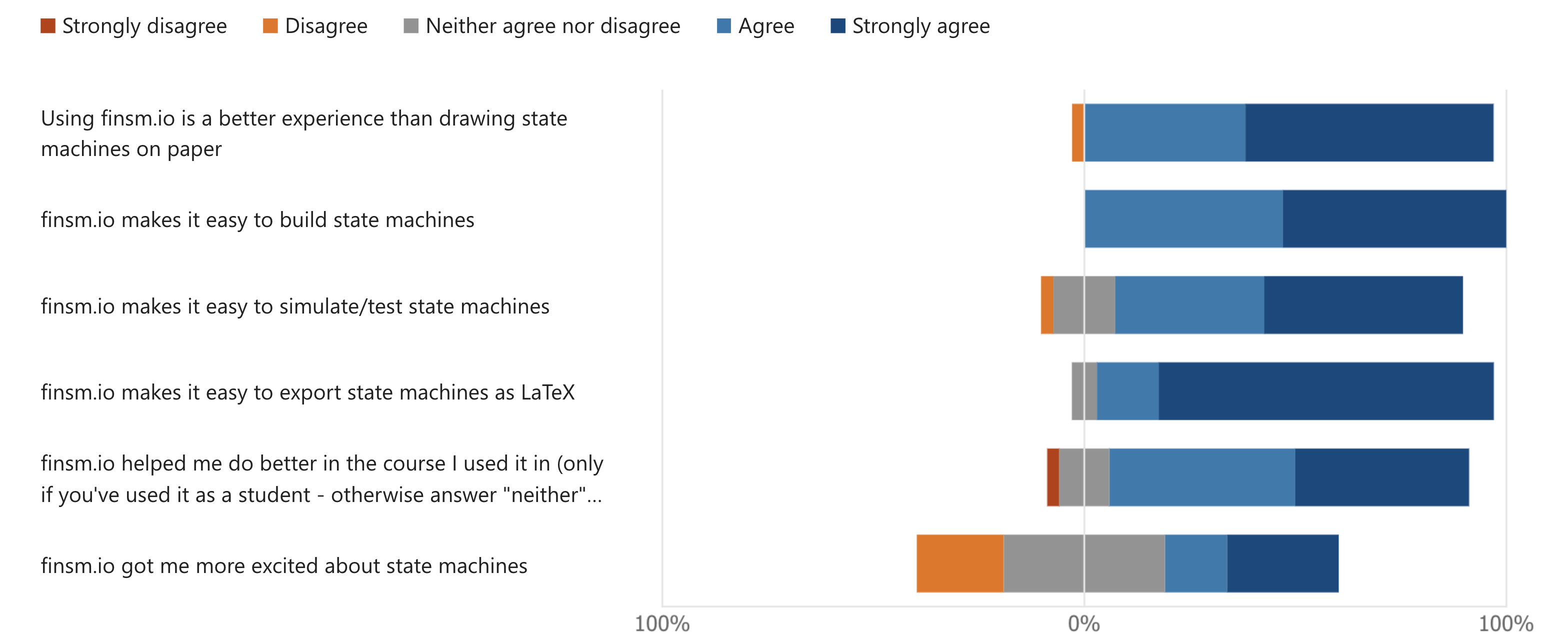}
    \caption{Likert scale questions asked of students [n = 34].}
    \label{Fig:Q3}
\end{figure*}

Figure~\ref{Fig:Q3} shows a Likert scale question asking them to rate their agreement with
several statements. Our tool scored highly for being a better experience than drawing on 
paper, ease of use, and many students agreed it helped them do better in the course they
were taking. Results were more mixed on whether it made them more excited to learn about
finite state machines.

Students were also asked to give written feedback. Most were positive, including one student
who said the experience was better than other general-purpose drawing tools. Some 
constructive feedback included that some actions were not intuitive when 
starting, a lack of useful keybinds, and mobile support. Echoing
the instructor and TAs, one student asked for support for PDA and Turing Machines, and expressed disappointment in having to draw those
by hand when it came to that part of the course.

\subsection{Threats to Validity}
From the instructor and TA perspective, only one instructor and two TAs were able to be 
reached for comment. Their perspective may also be skewed by virtue of knowing the creators
in an academic capacity.

On the student perspective, some students surveyed 
had not used the tool in a while, which could affect their judgment. This also means that
the most motivated users were more likely to answer the survey, likely skewing 
the answers towards positivity. To combat this in the future, we can build quick 
survey questions into the tool itself.

\section{Related Work}

The use of animations to teach concepts in computer science
curriculums is a wide topic of interest. 
In an algorithms course,
\citet{lawrenceEmpiricallyEvaluatingUse1994}
investigates the use of animations to teach algorithms
in classrooms, which led to higher accuracy on a post-test examination of understanding.
% textbf{CKA:  Were the animations of DFAs or something else?}
In class-based settings, \citet{berqueTeachingTheoryComputation2001}
describes a classroom equipped with pen-based computers, a touch-sensitive
electronic whiteboard and locally written groupware to teach a Theory
of Computation course. The survey results suggest that students prefer their method of course delivery over a traditional classroom setting. However, implementing this approach would require a specialized classroom environment or software that must be fully integrated and used consistently throughout the entire course.

\citet{pillayLearningDifficultiesExperienced2010} reports
a study conducted to identify learning difficulties on formal languages
and automata theory (FLAT).
% \textbf{CKA:  Can you say anthing about what the learning difficulties were?  Do they relate to visualization?}
They report that the main difficulty students encounter is a lack
of problem-solving skills rather than a lack of conceptualization or knowledge.
In this regard, the simulation mode of \texttt{finsm.io} complements
the problem-solving process by visualizing the states of the machine
for test inputs and provides helpful errors for incorrectly constructed machines.
Other similar tools exist, such as 
JFLAP by~\cite{rodger2006jflap, gramondUsingJFLAPInteract1999}, 
jFAST by~\cite{whiteJFASTJavaFinite2006}, and
ComVIS by~\cite{jovanovicTeachingConceptsRelated2021}. These 
tools were developed using the Java programming language.
\citet{barwise1994turing} presents a program developed
for the Macintosh. Mobile platforms have also been targeted,
including by \citet{singhAutomataSimulatorMobile2019}
and \citet{pereiraMobileAppTeaching2018}.
Of these tools, JFLAP is the most widely used, with Google Analytics
reporting a total of 714,535 new users from September 2012 to
May 2022 on their website. 
% \footnote{https://www.jflap.org}.
It also supports the most features. 
Compared to JFLAP, \texttt{finsm.io} is better adapted to today's
learning environment, as it is web-based, requiring
no installation, its simple interface avoids overwhelming new users, and \LaTeX\ output supports assignment workflows.
We believe that it is the combination of these
three features that has led to its adoption in the finite
automata courses at McMaster University, as previously a traditional
classroom approach was preferred by the instructors,
but they have since integrated \texttt{finsm.io}
into their courses.

Another approach taken, as opposed to working
with graphical representations of finite state machines,
would be to represent them with a domain-specific language (DSL).
\citet{chakraborty2007language} presents a
Turing Machine Description Language (TDML) for modeling Turing machines.
\citet{chakrabortyCompilerbasedToolkitTeach2013} uses the
Finite Automaton Description Language (FADL), with a toolkit
designed around the language to model and simulate finite automata.

The Elm programming language was developed as a functional approach to web programming by  \citet{czaplicki2012elm}.
It has been used at Brilliant.org, a STEM-education company, who used it to create Diagrammar%
\footnote{\scriptsize\url{https://www.thestrangeloop.com/2022/diagrammar-simply-make-interactive-diagrams.html}}
to produce interactive educational diagrams. 
It has been used for other academic diagramming tools, in the context of Model-Driven-Development (MDD):
\cite{schankula2020newyouthhack} created 
{PAL Draw} for specifying  client-server applications; and
\cite{pasupathi2022teaching} developed {SD Draw} for drawing state diagrams.
Neither of these MDD tools supports simulation or \LaTeX~output.

\section{Conclusions and Future Work} % (Both)
The tool presented shows promise for instructors teaching
courses involving finite state automata, which students agree helps with learning
and instructors agree helps facilitate teaching. 
It fits into the existing learning workflow by generating \LaTeX{} output.
Built using the functional programming 
language Elm, it is reliable and accessible as a free web app.

Future work on the tool will include implementing 
Push-Down Automata (for which there is already an active pull request) and Turing
Machines and adding more visualizations for common FSA algorithms. A controlled trial should seek to quantitatively show the effectiveness of the tool for teaching and 
learning. Finally, a submission system, automatic test-case marking system, and mark export to 
common learning management systems should be implemented. The authors welcome
contact from any interested instructors.

\section*{Acknowledgments}
Thanks to the instructors, TAs and students for their feedback. Thank you to Drs.~Christopher Anand and Spencer Smith 
for help ideating and revising this work.

\bibliographystyle{IEEEtranN}
\clearpage
\bibliography{BibRefs}
\end{document}